\documentclass[fleqn,usenatbib]{mnras}

\usepackage{newtxtext,newtxmath}
\usepackage{xcolor}

\newcommand\lsim{\mathrel{\rlap{\lower4pt\hbox{\hskip1pt$\sim$}}
\raise1pt\hbox{$<$}}}

\usepackage[T1]{fontenc}

\DeclareRobustCommand{\VAN}[3]{#2}
\let\VANthebibliography\thebibliography
\def\thebibliography{\DeclareRobustCommand{\VAN}[3]{##3}\VANthebibliography}

\usepackage{graphicx}	
\usepackage{amsmath}	

\title[A Multi-Wavelength View of RZ2109]{An 
Extreme Ultra-Compact X-ray Binary in a Globular Cluster: Multi-Wavelength Observations of RZ2109 Explored in a Triple System Framework}

\author[K. C. Dage et al.]{
Kristen C. Dage,$^{1,2}$\thanks{E-mail: kcdage@wayne.edu}\thanks{NASA Einstein Fellow},   Arash Bahramian, $^{3}$ Smadar Naoz, $^{4,5}$
 Alexey Bobrick,$^{6}$ Wasundara Athukoralalage,$^{7,8}$   \newauthor McKinley C. Brumback, $^{9}$  Daryl Haggard,$^{2}$ Arunav Kundu, $^{10}$ Stephen E. Zepf $^{7}$
\\
$^{1}$Wayne State University, Department of Physics \& Astronomy, 666 W Hancock St, Detroit, MI 48201, USA \\
$^{2}$Department of Physics, McGill University, 3600 University Street, Montr\'eal, QC H3A 2T8, Canada\\
$^{3}$International Centre for Radio Astronomy Research $-$ Curtin University, GPO Box U1987, Perth, WA 6845, Australia\\
$^{4}$ Department of Physics and Astronomy, University of California, Los
Angeles, CA 90095, USA \\
$^{5}$ Mani L. Bhaumik Institute for Theoretical Physics, Department of
Physics and Astronomy, UCLA, Los Angeles, CA 90095, USA \\
$^{6}$Technion - Israel Institute of Technology, Physics Department, Haifa 32000, Israel  \\
$^{7}$Department  of  Physics  and  Astronomy,  Michigan  State  University,  East Lansing, MI 48824, USA \\
$^{8}$ Center for Astrophysics | Harvard \& Smithsonian, 60 Garden Street, Cambridge, MA 02138-1516, USA \\
$^{9}$ Department of Physics, Middlebury College, Middlebury, VT 05753, USA \\
$^{10}$ Department of Physics, Birla Institute of Technology \& Science, Pilani, K K Birla Goa Campus, NH17 B, Zuarinagar, Goa 403726, India \\
}
\date{Accepted XXX. Received YYY; in original form ZZZ}

\pubyear{2023}

\begin{document}
\label{firstpage}
\pagerange{\pageref{firstpage}--\pageref{lastpage}}
\maketitle

\begin{abstract}
The globular cluster ultraluminous X-ray source, RZ2109, is a complex and unique system which has been detected at X-ray, ultra-violet, and optical wavelengths.  Based on almost 20 years of \textit{Chandra} 
 and \textit{XMM-Newton} observations, the X-ray luminosity exhibits order-of-magnitude variability, with the peak flux lasting on the order of a few hours. We perform robust time series analysis on the archival X-ray observations and find that this variability is periodic on a timescale of 1.3 $\pm 0.04$ days. The source also demonstrates broad [OIII] 5007 \AA\hspace{0.1cm}emission, which has been observed since 2004,  suggesting a white dwarf donor and therefore an ultra-compact X-ray binary. We present new spectra from 2020 and 2022, marking eighteen years of observed [OIII] emission from this source. Meanwhile, we find that the globular cluster counterpart is unusually bright in the NUV/UVW2 band.  Finally, we discuss RZ2109 in the context of the eccentric Kozai Lidov mechanism and show that the observed 1.3 day periodicity can be used to place constraints on the tertiary configuration, ranging from 20 minutes (for a 0.1 ${\rm M}_\odot$ companion) to approximately 95 minutes (for a 1 ${\rm M}_\odot$ companion), if the eccentric Kozai Lidov mechanism is at the origin of the periodic variability. 
\end{abstract}




\section{Introduction}

Ultraluminous X-ray sources (ULXs) are off-nuclear objects with X-ray luminosities exceeding the nominal Eddington limit of $10^{39}$ erg/s for an $\sim10\,{\rm M}_\odot$ black hole. 
The complex and unique ULX 
XMMU122939.7+075733, hereafter RZ2109, is located in an extragalactic globular cluster associated with the Virgo elliptical galaxy NGC 4472 (D=16.8 Mpc). RZ2109 is the first ever black hole candidate in a globular cluster \citep{2007Natur.445..183M}, and displays highly variable behaviour in both X-ray and optical bands \citep[and references therein]{Dage19b}.

One interpretation for RZ2109 is that it is a stellar-mass compact object accreting from an oxygen rich white dwarf companion, since the optical spectrum reveals one-of-a-kind broad [OIII] $\lambda$ 5007\AA\hspace{0.1cm}and 4959 \AA\hspace{0.1cm}emission lines with no excess hydrogen emission above the globular cluster continuum \citep{2008ApJ...683L.139Z,2014ApJ...785..147S}. \citet{2012ApJ...752...90P} find that the X-ray emission in the system is not significantly beamed, and analysis by \citet{2012ApJ...759..126P} and  \citet{Dage19b} determine that an ionised oxygen nebula with a size scale of up to 2 parsecs is present in the system, possibly related to mass loss from the system.

Systems where a compact object accretes from a white dwarf companion are known as ultra-compact X-ray binaries (UCXBs) because the accretion disk is expected to be spatially small, and the binary quite tight (with orbits less than 80 minutes; 
\citealt[and references therein]{2012A&A...537A.104V}). About half of all UCXBs are found in globular clusters \citep{Avakyan2023,2023arXiv230507691A}, but RZ2109 is the brightest known UCXB, and the only extragalactic one so far. \citet{Bildsten2004} suggest that most ULX sources in GCs of elliptical galaxies may also be young UCXBs. 

Binaries containing a white dwarf donor and a compact object (neutron star or black hole) can spiral towards coalescence due to gravitational wave emission. If the white dwarf mass is sufficiently low, the system will undergo stable mass transfer, producing a UCXB, e.g. \citet{2012A&A...537A.104V}. For neutron star accretors, UCXBs may only form from white dwarfs less massive than $0.2$~--~$0.3\,{\rm M}_\odot$ \citep{2017MNRAS.467.3556B}. For binaries containing black holes, even the most massive white dwarfs, depending on the black hole mass, may produce UCXBs \citep{2017MNRAS.467.3556B, 2017ApJ...851L...4C}. In contrast, if the white dwarfs are more massive than these limits, mass transfer becomes unstable, leading to a disruption of the white dwarf on a dynamical timescale and likely producing a faint supernova-like transient, e.g. \citet{Zenati2020,Bobrick2022}. Early on in their lives, UCXBs from white dwarf-neutron star/black hole binaries experience a phase of super-Eddington accretion and can be observed as ULXs, e.g. \citet{Bildsten2004, 2017MNRAS.467.3556B}. As per \cite{2017MNRAS.467.3556B}, given that massive carbon-oxygen or oxygen-neon WDs can only form ULXs with black hole companions, RZ2109 most certainly contains a black hole.

The UCXB 47 Tuc X-9,  with its measured 28 minute orbital period \citep{2017MNRAS.467.2199B}, is likely a Galactic relative of RZ2109 \citep[see, e.g., theoretical work by][]{2017ApJ...851L...4C}. Although 47 Tuc X-9 is relatively fainter in X-ray than RZ2109 (peak L$_x$= 8.4 $\times 10^{33}$ erg/s, \citealt{2017MNRAS.467.2199B}) with radio luminosity of 5.8 $\times$ $10^{27}$ erg/s, \citep{2015MNRAS.453.3918M}, it shows other similarities to RZ2109 including the presence of oxygen emission and a lack of excess hydrogen emission \citep[and references therein]{2017MNRAS.467.2199B, 2018MNRAS.476.1889T}. Thus, 47 Tuc X-9 offers a valuable comparison system for RZ2109, for example, as demonstrated by \cite{2023LRR....26....2A}, RZ2109's orders of magnitude higher mass transfer rate relative to that of 47 Tuc X-9 corresponds to an inferred orbital period close to 5 minutes.  Due to its ultra-compact nature, 47 Tuc X-9 is FUV bright, with an observed carbon emission line \citep{2008ApJ...683.1006K}. While it is not possible yet with current facilities to detect RZ2109 in radio, it may be possible to do so in the future with the ngVLA\footnote{\url{https://ngvla.nrao.edu/}}. However, the FUV bright nature of 47 Tuc X-9 motivates us to search existing UV data of RZ2109 to determine if there are other similarities.

As of now, no ULXs have been observed in Galactic globular clusters,  \citep[e.g.,][]{Avakyan2023}, but to date a number of ULXs have been characterised in over 20 extragalactic globular clusters thanks to decades of \textit{Chandra} and \textit{XMM-Newton} observations \citep[e.g.][]{Dage19a}. Both the X-ray and optical wavelength can be employed to characterise the optical properties of the host clusters, search for optical emission (such as in the case of RZ2109), and probe the X-ray spectra and X-ray variability. Although the current number of ULXs in globular clusters that have been studied in this manner is small, the sample is diverse in terms of best-fit X-ray spectral parameters and short and long-term X-ray variability \citep{Dage2020}. 

Two of these sources in particular have been heavily observed since their discovery. \citet{Irwin} observed the presence of narrow [NII] and [OIII] emission lines in the NGC 1399 source GCU7; that in combination with the high X-ray emission was suggested to be an intermediate mass black hole tidally disrupting a horizontal branch star in the globular cluster \citep{Irwin,2012MNRAS.424.1268C}. Horizontal branch stars are present in globular clusters with multiple stellar populations; and thus their far ultra-violet luminosities will be enhanced by helium burning from extreme horizontal branch stars \citep{2009Ap&SS.320..261C,2012AJ....144..126D}. An alternate explanation by \citet{2011MNRAS.410L..32M} is that the source is a compact object accreting from an R Corona Borealis star. 

A number of explanations have also been suggested for RZ2109; however, the longevity of its X-rays as well as its high X-ray variability rules out a tidal disruption event of a white dwarf by an intermediate-mass black hole \citep{2011ApJ...726...34C}. \citet{2012MNRAS.423.1144R} suggested that the oxygen emission may be explained by a nova shock ejecta serendipitous with a luminous X-ray source in the globular cluster; this has been ruled out by the longevity of the X-rays \citep{2011ApJ...739...95S, Dage18}. \citet{Maccarone10} suggested that one possible explanation for this source is a hierarchical triple system with a black hole/white dwarf binary.  Although several studies have suggested that RZ2109 has an intermediate-mass black hole primary \citep{2010MNRAS.407L..59P, stiele, tiengo}, this interpretation is at odds with the shape of the broad [OIII] emission line (over 1500 km/s, with a narrow Gaussian core and double-peaked emission, \citealt{zepf08}) which rules out models like that of \citet{2010MNRAS.407L..59P} that suggest the origin of the [OIII] emission line could be due to rotation from a massive black hole \citep{2011ApJ...739...95S}. 

As suggested by \cite{tiengo}, the X-ray variability from RZ2109 may bear some relation to that seen in quasi-periodic eruptions (QPEs); strong X-ray variability in the soft band, and in the case of the QPE discovered in GSN 069, lasted for approximately an hour, recurring every 9 hours for a total of 54 days \citep{2019Natur.573..381M, tiengo}. RZ2109 is also a soft source, and the order of magnitude variability lasts for several hours \citep{2007Natur.445..183M}, and has been observed several times in 16 years worth of \textit{Chandra} and \textit{XMM-Newton} observations, although we note that the spectrum of RZ2109 is distinctly different from those of QPEs, e.g. GSN 069 which is super-soft \citep{2023A&A...670A..93M}.

\begin{figure*}
    \centering
    \includegraphics[width=6in]{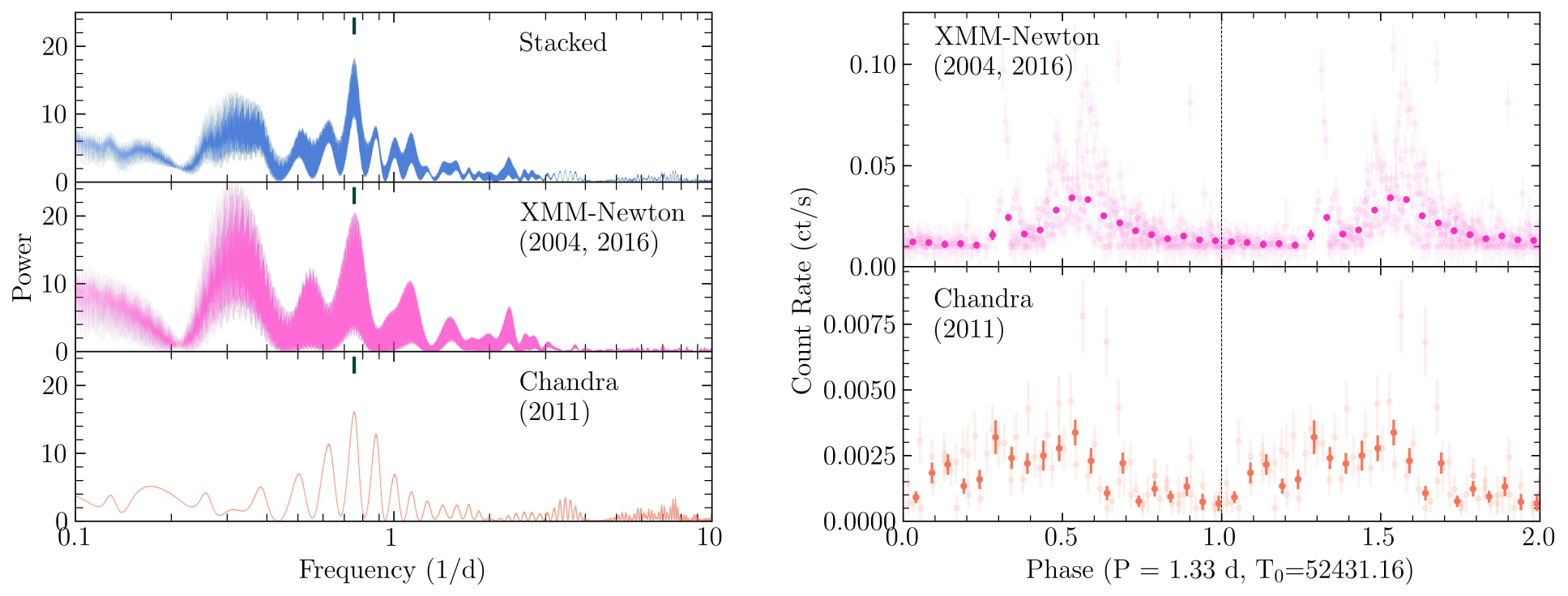}
    \caption{(Left) Individual \textit{Chandra}, \textit{XMM-Newton} and their stacked MHAOV Periodograms. The black vertical bar indicates the peak-power period of 1.33 day. (Right): Phase-folded lightcurves from \textit{XMM-Newton} and \textit{Chandra} using the most-likely period found by periodogram analysis. Faint data points represent light curves binned by 2 ks (i.e., unbinned in phase). $T_0$ indicates the chosen zero-phase point in MJD for folding all light curves.}
    \label{fig:xraylc}
\end{figure*}

RZ2109's spectral evolution is also unique: \citealt{Dage18} report that it is best fit by a black body disc and power-law model, with the inner disc temperature quite soft (around 0.12 keV). The best-fit inner disc temperature does not vary with luminosity, which may be consistent with obscuration by disc winds, a scenario suggested by \citealt{Maccarone10}. 

Here, we suggest that the long-term periodicity observed in RZ2109 may be explained by torques induced via a third body in the system orbiting at a large distance from the centre of mass, otherwise known as the Eccentric Kozai-Lidov (EKL) mechanism \citep[e.g.,][see the latter for a review]{Kozai,Lidov,Naoz16}. In this picture, RZ2109 is composed of a tight inner binary, and the third object on a stable orbit induces eccentricity and inclination oscillation on secular (long-term) timescales. This type of hierarchical configuration allows for the utilization of the secular approximation (phase averaged, long-term interaction), where the interactions between RZ2109 and the third body are equivalent to two orbits that are treated as massive ``wires,'' \citep[see for review][]{Naoz16}. The eccentricity and inclination oscillation occurs on a characteristic timescale, which is a function of the two orbital periods, the outer orbit eccentricity and masses \citep[e.g.,][]{Antognini15}. Below, we constrain the possible eccentricity of the third body and period by identifying RZ2109's long-term periodicity with EKL oscillations.

We analyse X-ray, UV and optical observations of RZ2109. The observations and data analysis methods are presented in Section \ref{sec:data}, the results are discussed in Section \ref{sec:res}, and our conclusions and recommendations for  future follow-up are summarized in Section \ref{sec:summ}.

\section{Data and Analysis}
\label{sec:data}

The globular cluster ultraluminous X-ray source RZ2109 has been observed at many different wavelengths with a variety of instruments. These include observations of RZ2109 in X-ray for over 30 years with the Roentgen Satellit (ROSAT, \citealt{1982AdSpR...2d.241T}), the \textit{Chandra} X-ray Observatory \citep{2000SPIE.4012....2W}, the X-ray Multi-Mirror Mission (XMM-Newton, \citealt{2001A&A...365L..18S}), and the Neil Gehrels Swift Observatory \citep{2004ApJ...611.1005G}, in optical with the Gemini-South Observatory and the Southern Astrophysical Research (SOAR) telescope, and in the near ultra-violet with the Neil Gehrels Swift Observatory's Ultra-violet Optical Telescope (UVOT). 

In this work we specifically undertake statistically robust timing analysis of over 20 years worth of X-ray lightcurves obtained by \textit{Chandra} and XMM-Newton to search for any periodicity in the previously noted extreme variability on the time scale of hours \citep{2007Natur.445..183M}. We also utilise 16 ks of observations from Swift's UVOT instrument, taken in 2008, 2010, 2019, and 2020 with the UVW2 (NUV) filter, as well as new optical spectroscopy from SOAR obtained in 2020, and 2022.  

\subsection{X-Ray Observations}
RZ2109 has been observed in X-ray as early as 1994 \citep{1996ApJ...471..683I}. Since 2000, it has been observed 14 times by \textit{Chandra} and 7 times by XMM-Newton. As noted by \citet{stiele,tiengo}, the $\sim$ hourly variability shows suggestions of a periodic nature. 

We reprocessed the \textit{Chandra} observations using \textsc{ciao} command \texttt{repro}. The \textit{Chandra} background-subtracted lightcurves were extracted in the 0.2-10 keV band, using \textsc{ciao} and binned to 100 seconds, with 3 arcsecond source regions. For the XMM-Newton observations, we extracted the background-subtracted lightcurves with the \textsc{XMM-SAS} command \texttt{evselect}, from EPIC-pn CCDs only, using a 45 arcsec extraction radius, in the 0.2-10 keV band, PATTERN<=4, FLAG== 0 and binned to 100 seconds. 

 We performed timing analysis on all of the longer data sets: four of the XMM-Newton observations (one in 2004 and three observations in 2016 spaced two days apart each) and two of the \textit{Chandra} observations (two observations in 2012 spaced one week apart). As discussed below, many of these lightcurves have been contaminated due to background flaring, which complicates the timing analysis.

\begin{table}
\centering
\caption{\label{table:chandraobsinfo}  \textit{Chandra} and \textit{XMM-Newton} observation log. Each of these long observations contains one full flare.} 
\begin{tabular}{lccccc}
\hline
\hline
ObsID     &Date  & Exposure\\ 
&& (seconds) \\
\hline

12889 (PI: Kraft)  & 2011-02-14 &  135590	 \\ 
12888  (PI: Kraft) & 2011-02-21 & 159310   \\

\hline
0200130101 (PI: Maccarone)   & 2004-01-01  & 91926 \\ 
0761630101 (PI: Maccarone) & 2016-01-05 &  115041  \\
0761630201 (PI: Maccarone)&2016-01-07& 112407 \\
0761630301 (PI: Maccarone) &2016-01-09& 114063\\

\hline
\end{tabular}
\end{table}

\subsubsection{Accounting for background flares in observations} 
We performed timing analysis on six long observations total (Table \ref{table:chandraobsinfo}). As discussed in \citet{Dage18}, most of the XMM-Newton observations suffered from background flares. ObsID 0761630201 was the most affected by background flaring, with three background flares occurring during the observation. For the XMM data, we used the GTIs from \citet{2007Natur.445..183M} to clean the 2004 observation and analysis from \citet{stiele} for the set of 2016 observations. After analysing the background lightcurve of Chandra ObsID 12889, we removed data from 80 ksec onward as that was contaminated by a background flare.

\subsubsection{Timing analysis}

We searched for periodicity by employing the phase dispersion minimization analysis package \textsc{p4j},  \citep{huijse}. Due to the sensitivity differences in the detectors of \textit{Chandra} and \textit{XMM-Newton}, 
We first extracted light curves and periodograms for each instrument (pn for \textit{XMM-Newton}, ACIS for \textit{Chandra}) separately. We implemented the Multi-harmonic Analysis of Variance \citep[MHAOV,][]{SchwarzenbergCzerny1996} periodogram, and searched for periodicity with frequency between 0.1 and 10 day$^{-1}$, with a search resolution of $10^{-4}$ day$^{-1}$. As detailed in \citet{huijse}, no particular model or underlying probability density is assumed for the light-curve, and the method is generally more robust against artefacts due to noise and sparse sampling than many other alternatives \citep{Graham2013}. 

We then stacked the power-spectra from the two different observatories to find the most likely period, which is $1.33\pm0.04$ days. We perform bootstrap sampling following \citet{huijse} to assess significance of this possible periodicity. We find null-hypothesis probabilities\footnote{Our null hypothesis is that the peak in the periodogram is produced by random uncorrelated noise, under sampling conditions similar to the datasets under investigation.} of 2\% and 41\% for \textit{Chandra} and \textit{XMM-Newton} light curves respectively. While a bump at a period of 1.33 day present in both \textit{Chandra} and \textit{XMM-Newton} datasets, it is not significant in the \textit{XMM-Newton} light curve, where we estimate a probability of 41\%, and thus we cannot rule out the null hypothesis when considering the \textit{XMM-Newton} data alone. It is the detection in the \textit{Chandra} light curve that draws attention to this signal. Following Fisher's combined probabilities test \citep[e.g.,][]{Fisher1992} to account for the number of trials, combining the null hypothesis probabilities yields a fused probability of 4\% for the null hypothesis.

We note that the data are very sparsely and unevenly sampled (years apart). However, performing this analysis on data from different epochs and X-ray instruments yields remarkably similar measurements for a periodic signal (with similar phase-resolved behavior from a common zero-phase point) in the X-rays. It is thus suggestive that the flaring behaviour seen in RZ2109 is indeed an intrinsic property of the source, and likely periodic. The individual and stacked power-spectra, and folded lightcurves are displayed in Figure \ref{fig:xraylc}.


\subsection{Ultra-Violet Observations}
Swift/UVOT has observed RZ2109 with the NUV UVW2 filter (central wavelength 1928 Angstrom) for over 16 ks from 2007 to 2020 (Table \ref{table:uvot}). 
\begin{table}
\label{table:uvot}
\caption{UVOT observation log}
\begin{tabular}{lll}
 \hline
ObsID       & Date       & Exposure Length\\
&& (seconds) \\ \hline  \hline
00036575001 & 2007-11-13 & 4360 \\
00031078001 & 2007-12-25 & 1815            \\
00031078002 & 2007-12-27 & 1899            \\
00031078003 & 2010-03-22 & 2168            \\
00031078004 & 2010-03-26 & 1805            \\
00031078005 & 2010-03-30 & 2087            \\
00031078006 & 2019-04-15 & 2407            \\
00031078007 & 2019-05-13 & 1048            \\
00031078008 & 2019-05-17 & 886             \\
00031078009 & 2019-05-27 & 1297            \\
00031078010 & 2019-05-31 & 285             \\
00031078011 & 2019-06-05  & 785            \\
00013603001 & 2020-07-03 & 1170            \\
00013603002 & 2020-07-08 & 1118            \\
00013603003 & 2020-07-13 & 1226            \\\hline
\end{tabular}
\end{table}

Using Swift analysis tools, \textsc{uvotimsum} we stacked all existing UVOT/UVW2 images and exposure maps, for a total of 16553.81 seconds. We used \textsc{uvotsource} to measure the UVW2 magnitude of RZ2109 with a 3.0" aperture, manually setting a background region of the same size. It was detected at 5.9 $\sigma$ with a Vega magnitude of 22.14 $\pm$ 0.20. 

\subsubsection{Mitigating the Effect of Red Leak in UV detectors}

Red leak is a concern with many current UV detectors; as optical light from redder objects can be picked up as UV flux \footnote{\url{https://swift.gsfc.nasa.gov/analysis/uvot_digest/redleak.html}}.  By measuring the flux in the stacked images of nearby confirmed GCs, we can determine if red leak is adding a baseline UV detection to all the sources overall. Using a spectroscopic catalogue of NGC 4472 globular clusters from Bergond et al (in prep), we performed photometry on any spectroscopically selected GCs which fell on at least 8ks of stacked data. We performed the photometry the exact same way as with RZ2109, with \textsc{uvotsource} using 3.0" apertures with manually created background regions. We removed young star clusters identified in \citet{Battaia12} which are a product of a recent merger of NGC 4472 with VCC 1249. 

As seen in Figure \ref{fig:UV}, nine other globular clusters besides RZ2109 were detected above 3 $\sigma$, 17 were marginal detections, and 32 were non-detections. This suggests that the FUV-enhanced signal detected in RZ2109 is intrinsic to the source, and not solely a product of red leak.

\subsection{Optical Spectroscopy}

\begin{figure}
    \centering
    \includegraphics[width=3.5in]{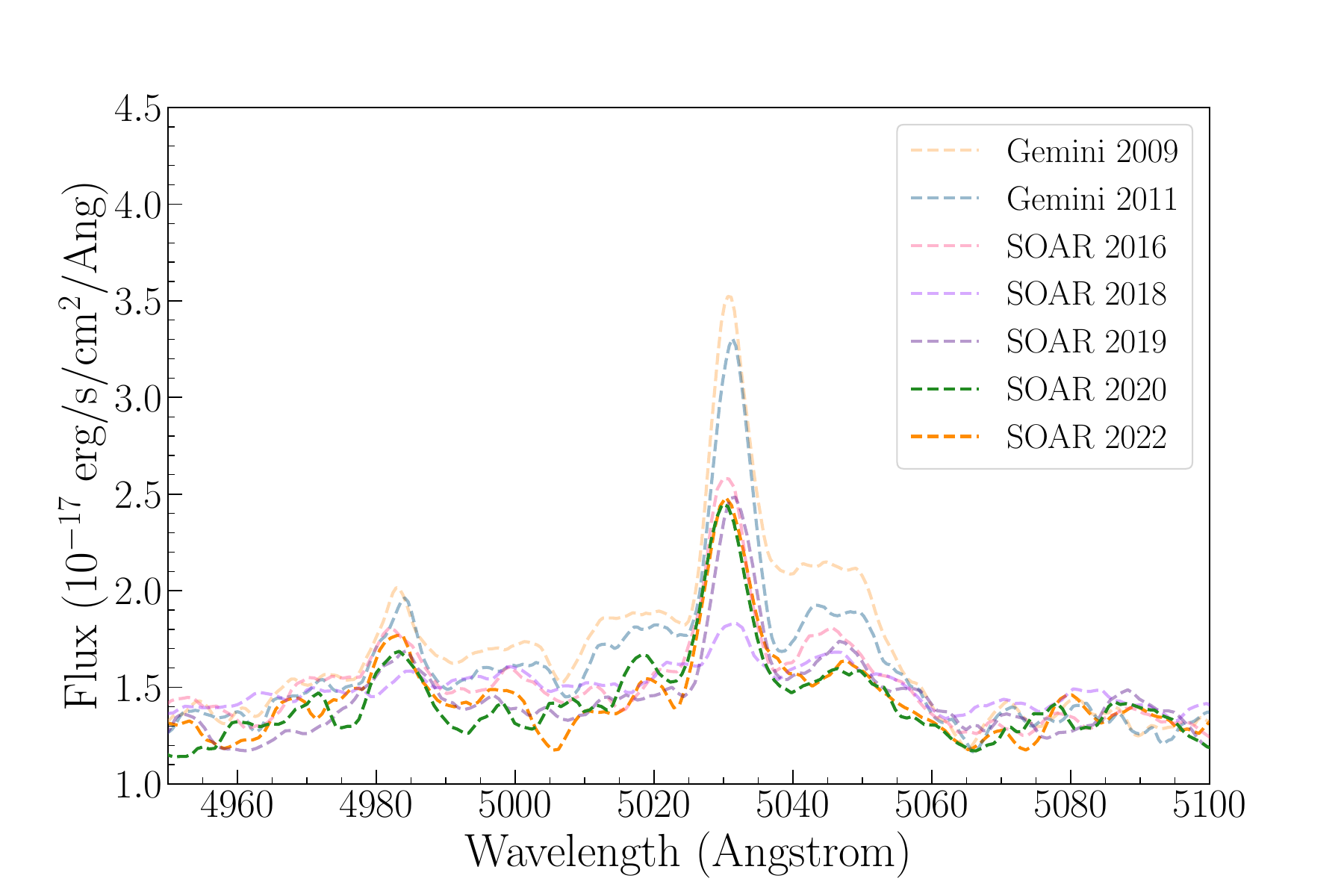}
    \caption{Smoothed optical spectra of RZ2109's [OIII] emission line. The two new spectra from 2020 and 2022 are presented alongside the 2009, 2011, 2016, 2018, and 2019 spectra from \citet{Dage19b}. The optical emission line does not appear to have changed significantly in the new observations, and the equivalent widths are the same as in 2019 (within measurement uncertainties). }
    \label{fig:optspeectra}
\end{figure}

RZ2109's broad and unique [OIII] 5007\(\text{\AA}\), 4959\(\text{\AA}\) emission line has been observed by many telescopes beginning in 2004 \citep{2004AJ....127..302R}. \citet{2007ApJ...669L..69Z,zepf08, 2011ApJ...739...95S, 2014ApJ...785..147S, 2012ApJ...752...90P, 2012ApJ...759..126P, Dage19b} all discuss the evolution of this emission line and implications for the nature of the source. We have further observed the emission line with SOAR in 2020 and 2022, using the Goodman High Throughput Spectrograph red camera with the 900 l/mm grating centred at 495 nm and the 1.2" longslit. We used IRAF to extract the spectra; the spectrum was traced and optimally extracted with \texttt{apall}, corresponding arc spectra were extracted with \texttt{apall}. We determined the arc solution and wavelength calibrated with \texttt{identify}. The emission line was observed in all instances, and using the same techniques as \citet{Dage19b}, we normalize the 2020 and 2022 SOAR spectra, and measure the equivalent width in the 5007 \(\text{\AA}\) emission line. We find an equivalent width of 16.0 $\pm$ 1.3 in the 2020-02-25 observation, and 15.3  $\pm$  1.7 in the 2022-03-28 observation. The new spectra are displayed compared to the spectra from \citet{Dage19b} in Figure \ref{fig:optspeectra}.


\section{Results and Discussion}
\label{sec:res}

\begin{figure*}
    \centering
    \includegraphics[width=6.5in]{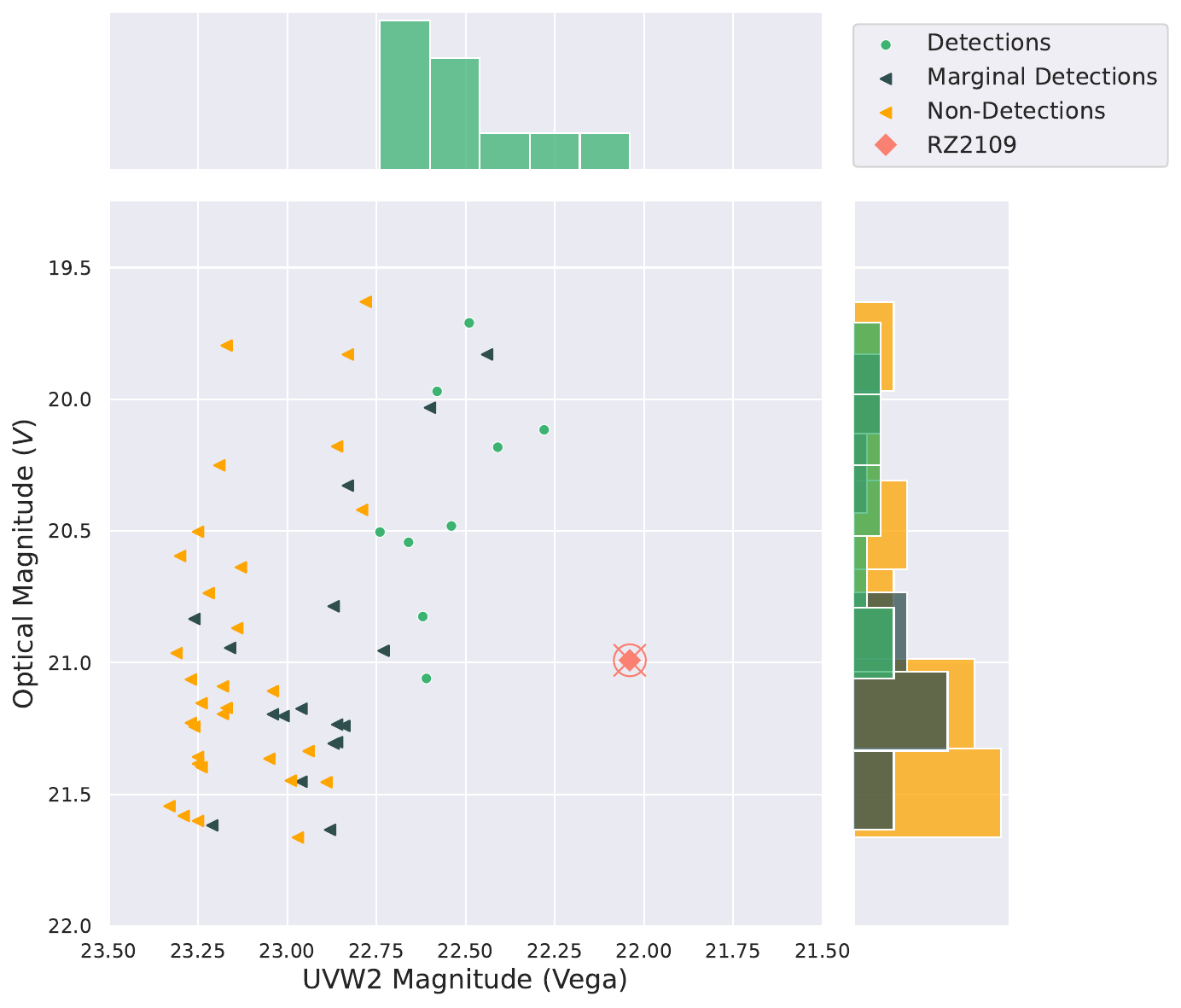}
    \caption{NUV magnitudes for spectroscopically confirmed GCs in NGC 4472, measured from stacked UVOT/UVW2 images from archival Swift data (2007 \& 2010), for a total of 16ks. While the distribution of optical magnitudes is well sampled, the majority of clusters are not detected above 3 $\sigma$. }
    \label{fig:UV}
\end{figure*}
We present X-ray timing, NUV photometry and optical spectroscopy of the globular cluster ultraluminous X-ray source RZ2109, using data obtained from \textit{Chandra}, XMM-Newton, Swift and SOAR. X-ray timing of Chandra and XMM-Newton observations suggests a 1.3-day recurring periodicity in the source, which last on average for $\sim$ 4.5 hours. Photometry of Swift NUV/UVW2 observations revealed that the host cluster is unusually bright. While red leak may play a role, clusters of similar optical magnitude to RZ2109 were not similarly detected, suggesting that red leak alone cannot explain the detection. Continued monitoring of RZ2109's broad [OIII] 4959\(\text{\AA}\), 5007\(\text{\AA}\) emission line in 2020 and 2022 shows that the flux increase seen by \citet{Dage19b} has stayed constant.  Below we discuss several scenarios which may explain the observations of this unique source. 

\subsection{X-ray Variability - a warped precessing accretion disc, or hierarchical triples?}
Although the recurring flares in RZ2109 have been suggested to be similar to those observed in quasi-periodic eruptions \citep{tiengo}, there are differences between the RZ2109 and QPE sources in both the X-ray spectral shape and also the observed periodicities. Specifically, the recurring QPE GSN 069 shows one hour flares which occur $\sim$ every nine hours in four out of eleven XMM-Newton observations spanning from 2010-2021 \citep{2023A&A...670A..93M}. In contrast, RZ2109's flaring behaviour is seen in long XMM-Newton and Chandra observations from 2004, 2011 and 2016, and the time series analysis we perform measures this period consistently across all of these data-sets. Thus, there is no evidence that RZ2109's behaviour is not strictly periodic. These observed differences the periodic behavior suggest that the mechanism that produces RZ2109's X-ray flares may be different than those behind QPE sources. Therefore we discuss two possibilities for a consistent mechanism to produce this variability: super-orbital period variability, and the extreme Kozai Lidov mechanism.

Super-orbital periodic variability is found in many types of X-ray binaries from neutron stars \citep{2003MNRAS.343.1213C} to black holes \citep{2022MNRAS.509.1062T}, and, notably including ULXs; \citep{2020MNRAS.495L.139T}. The observed super-orbital modulations can show orders-of-magnitude changes in flux \citep[e.g.][]{2003MNRAS.339..447C,2019MNRAS.482..337D}, and one possible interpretation of them is that of a warped, precessing accretion disc. RZ2109's Galactic analogue, 47 Tuc X-9 shows evidence for a 6.8 day order-of-magnitude modulation, in addition to its detected 28-minute orbital period \citep{2017MNRAS.467.2199B}.  One possible interpretation for RZ2109's variable nature is that of a precessing, warped accretion disc. 

On the other hand, due to the effects of a hierarchical triple, the outer companion can regulate the orbit of the mass-transferring donor by exchanging the donor's eccentricity by the inclination with respect to the tertiary.
As discussed in \citet{Maccarone10}, an ultra-compact system with a $<0.1 {\rm M}_\odot$ companion is a viable dynamically hard system that can form and survive in an old globular cluster such as the host of RZ2109. 

\begin{figure}
    \centering
    \includegraphics[width=\linewidth]{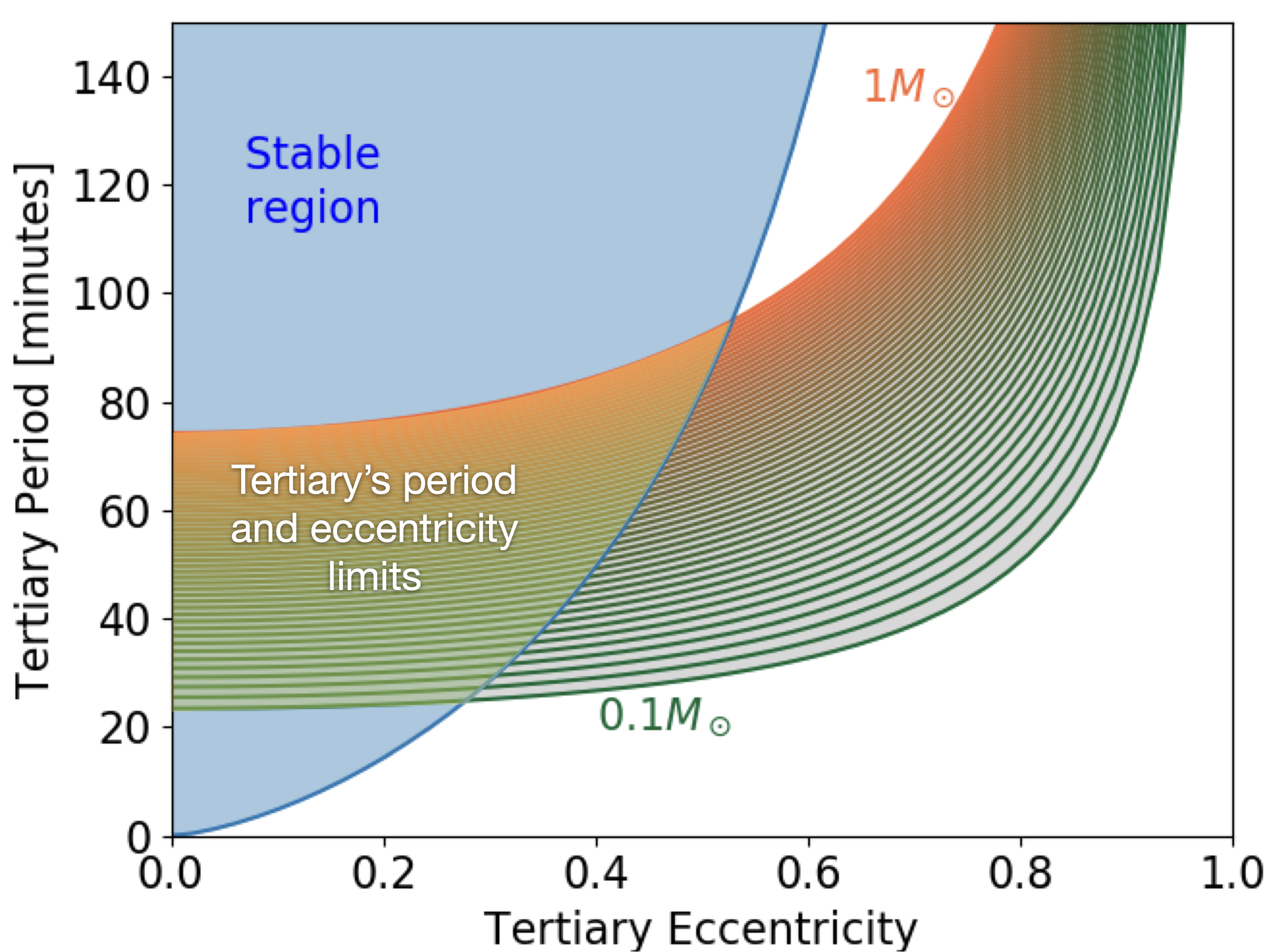}
    \caption{{ The possible tertiary's parameters. } 
    We consider the region at which a tertiary may form a stable configuration with the RZ2109 binary, based on having $\epsilon\leq0.1$, see Equation (\ref{eq:epsilon}), shown as the left shaded region. We also show the solution for the tertiary period as a function of its eccentricity, assuming that the EKL timescale corresponds to the observed periodicity. We adopt a black hole mass of $10$~M$_\odot$ and assume that the donor star has a negligible mass in comparison.  We consider tertiary's masses of $0.1$~M$_\odot$ (bottom, green line), up to $1$~M$\odot$, (top, red line). The intersection between the stable region and the EKL timescale are used to provide the constraints. For example, a $0.1$~M$_\odot$ star may have a period between $\sim 23-25$~minutes with eccentricity $\lsim 0.27$. This figure is agnostic to any formation channel and thus provides a robust estimation of the possible tertiary orbital configuration.  }
    \label{fig:tertiary}
\end{figure}

The EKL mechanism can result in eccentricity and inclination excitations on the BH-companion (the X-ray binary) orbit \citep[e.g.,][]{Naoz16}. In RZ2109, the observed $1.3$~days periodicity can be attributed to the inclination oscillations. Additionally, eccentricity excitations may trigger strong  episodes of accretion. The regularity of the observed $1.3$~days oscillations suggests that if this system has a third body, it results in a regular, non-chaotic quadrupole level of oscillations, and the octupole-level effects are suppressed \citep[e.g.,][]{Naoz+13sec,Li+14}.  In other words, the quadrupole level of the approximation timescale is roughly $1.3$~days. This timescale can be written as: 
\begin{equation}\label{eq:tEKL}
    t_{\rm EKL} \sim \frac{16}{30 \pi} \frac{m_{\rm BH}}{m_t}\frac{P_t^2}{P_{\rm RZ}}\left(1-e_t^2 \right)^{3/2} \sim 1.3~{\rm days}\ ,
\end{equation}
where $m_t$, $P_t$, and $e_t$ are the tertiary's mass, period, and eccentricity, respectively, and $P_{\rm RZ}=5$~minutes. 
In Figure \ref{fig:tertiary}, we solve for $P_t$ as a function of $e_t$, assuming that the EKL timescale is $1.3$~days. The bottom, solid black line represents $m_t=0.1$~M$_\odot$, while the top, dashed black line represents $m_t=1$~M$_\odot$.

Furthermore, the stability requirements of the hierarchical system may be expressed using the parameter $\epsilon\leq 0.1$, where:
\begin{equation}\label{eq:epsilon}
    \epsilon = \frac{a_{\rm RZ}}{a_t} \frac{e_t}{1-e_t^2} \ ,
\end{equation}
where $a_{\rm RZ}$ ($a_t$) is RZ2109’s (tertiary's) semi-major axis. This $\epsilon$ parameter is the prefactor of the octupole level of approximation in the Hamiltonian \citep[e.g.,][]{Naoz+13sec}. It was noted that this parameter is consistent with other stability criteria often used in the literature \citep[e.g.,][]{Naoz+14}, and we utilize it here for its simplicity. The blue-shaded area in Figure \ref{fig:tertiary} shows the parameter space of a Keplerian orbit that satisfy $\epsilon\leq 0.1$.

Thus, combining Equations (\ref{eq:tEKL}) and (\ref{eq:epsilon}), we find the possible orbital configuration a tertiary may have to induce the $1.3$~days periodicity. For example, a $0.1$~M$_\odot$ companion is constrained to have a period between $\sim 23-25$~minutes with eccentricity $\lsim 0.27$. We consider masses up to $1$~M$_\odot$, although given the age of the globular cluster \citep[$12$~Gyrs, e.g.,][]{2011ApJ...739...95S}, such a companion may possibly represent a white dwarf rather than a main sequence star. 

Figure \ref{fig:tertiary} does not depend on any formation channel. It only assumes that the observed brightness modulations are induced by a third, relatively far-away companion. The minimum timescale to induce orbital modulations on the inner orbit (the X-ray binary) is the EKL timescale, Eq.~(\ref{eq:tEKL}), which is used to find the dependency of the tertiary's period as a function of its eccentricity. Such a triple configuration requires stability\footnote{Recently, \citet{Zhang+23} demonstrated that a system that defies dynamical stability does not instantaneously ``break up'' (or dramatically alter the system's configuration). They found a well-defined timescale associated with stability deviation. This timescale is much smaller than the observational time (tens of years). Thus, since the observations show a stable X-ray binary with stable modulations, our interpretation of stability seems justified.   }, which we quantify using Equation  (\ref{eq:epsilon}). In other words, a triple configuration, regardless of its formation channel, has to satisfy these two relations, thus yielding a constraint on the third object's orbital configuration, ranging from 20 minutes for a 0.1 ${\rm M}_\odot$ companion to a maximum tertiary period of 95 minutes for a 1 ${\rm M}_\odot$ companion.

Note that a condition to have such a period excite eccentricity and inclination oscillation, the EKL timescale needs to be shorter than RZ2109's general relativity precession \citep[e.g.,][]{Naoz+13GR}. This precession for a circular orbit is about  $11.3$~days, longer than the EKL timescale.   

Lastly, we speculate on the formation channel of such a triple system. At face value, stellar evolution of triple stellar systems that produce X-ray binaries results in a wider configuration for the tertiary stars\citep[see figure 4 in][]{Naoz+16xrb}. However, supernova kicks that were neglected in previous studies may alter the configurations resulting in tighter orbital configurations \citep[e.g.,][]{Hamers18,Lu+19}. Moreover, kicks associated with white dwarf formation \citep[e.g.,][]{El-Badry+18} may also result in a tighter configuration in triples \citep[e.g.,][]{Shariat+23}. Furthermore, interaction with the surrounding gas may assist in forming a tight configuration \citep{Glanz+21,Hirai+22}, even in the substance of kicks. 
Thus, if RZ2109 is indeed a triple, it may have formed via a combination of EKL of a triple body stellar system involving a supernova kick, white dwarf kick, or orbital energy dissipation to result in this tight configuration.

Another possible formation channel may involve fly-by of wide binaries in the field \citep[e.g.,][]{Michaely+16,Michaely+20}. Particularly, a resonant encounter may lead to a dramatic orbital configuration change,
leaving a tight triple configuration \citep[e.g.,][]{Hills80,Sigurdsson+93,Samsing+17}. Recent developments in these types of chaotic encounters \citep[e.g.,][]{Stone+19,Ginat+21,Manwadkar+21,Kol23} may prove to be promising in constraining this possible channel in future studies.

\subsection{Near Ultra-Violet Signatures - X-ray/UV reprocessing or multiple stellar populations?}

A deep (16 ks) stacked Swift UVOT NUV image detected counterparts to a handful of GCs in NGC 4472, including a 5$\sigma$ emission detection of RZ2109 (Figure \ref{fig:UV}). Photometric measurements of RZ2109 in the Swift UVW2 band (similar to HST NUV) gives a Vega magnitude of 22.0. Even taking conservative estimates for the UV magnitudes observed for clusters \citep{Peacock17}, RZ2109 is at least a magnitude brighter than predicted.  As seen in Figure \ref{fig:UV}, RZ2109 is the only GC with an optical magnitude fainter than V=20.75 with significant NUV emission among a large number of clusters at similar optical magnitudes. Many other clusters of similar brightness were non-detections in the Swift UVW2 images, as well as some of the more massive clusters. 

This raises the question of whether the UV emission is associated with the X-ray binary or the host cluster. Observations of systems like SS433 \citep{2021MNRAS.506.1045M}, and other ULXs \citep{2023MNRAS.524.4302K}, suggest that X-ray/UV reprocessing is expected in extreme systems like ULXs. Reprocessing in the accretion disk of the ULX source in RZ2109 represents a likely origin for this excess UV emission, and since RZ2109's X-ray emission is known to vary significantly, it is expected that the UV emission also varies. 

However, it is possible that the excess NUV emission could be suggestive of multiple stellar populations in the globular cluster.  \citet{2023MNRAS.518...87D} note excess FUV emission in NGC 1399's globular cluster systems due to the presence of cool horizontal branch stars (which in turn may be the outcome of multiple stellar populations), which was not an indicator of the presence or luminosity of X-ray emission from the cluster. Multiple stellar populations may also be strongly correlated with the populations of X-ray binaries in globular clusters \citep{2022ApJ...931..149R}. However, we caution the evidence for multiple stellar populations can only be verified with FUV observations, as there may be other stars beyond the extreme horizontal branch stars emitting in NUV. Regardless, the extremely bright NUV detection is suggestive of something very unusual in the host cluster.


\section{Summary and Future Work}
\label{sec:summ}
We present analysis of X-ray, UV and optical spectroscopy of the globular cluster ULX RZ2109. Two new optical spectra from 2020 and 2022 show that the broad [OIII] emission lines are still observed, even after 18 years. Thanks to archival Swift/UVOT UVW2 observations, we find that the globular cluster host is significantly brighter in the UV band than nearby globular clusters, which suggests that either the ULX source is producing excess UV emission, or the host cluster is brighter in UV due to multiple stellar populations.  We perform a robust time series analysis of archival X-ray observations of the globular cluster ULX RZ2109, resulting in the measurement of a 1.3 day periodicity. If the 1.3 day periodicity is due to to the EKL mechanism, then we can constrain the tertiary configuration, which ranges from 20 minutes (for a 0.1 ${\rm M}_\odot$ companion) to 95 minutes (for a 1 ${\rm M}_\odot$ companion). 

The multiwavelength data presented here builds a solid foundation for determining the nature and future evolution of RZ2109 and provides instrumental information on the early evolution of UCXBs. 
The detection of oxygen emission lines and the observed oxygen nebula requires that RZ2109 is a black hole UCXB since neutron stars in binaries with carbon-oxygen white dwarfs experience unstable mass transfer \citep{2017MNRAS.467.3556B}. The X-ray luminosity constrains the mass transfer rate and hence the age of the system. Connecting the X-ray data to the optical data could lead to much more accurate estimates of the system's period and hence age \citep{2012A&A...543A.121V}. However, the specific relation between X-ray periodicity and optical appearance is non-trivial since the X-rays are reprocessed as they propagate, and any periodicity gets naturally smoothed away.  

UCXB theory \citep[e.g.,][]{2017MNRAS.467.3556B} \hspace{0.1mm} can further strengthen the argument that RZ2109 is a stellar mass black hole. Thus its presence in a globular cluster useful for comparisons of model predictions for the formation of stellar mass black hole - white dwarf binaries in star clusters. One channel for making such binaries is collisions between black holes and giant stars in globular clusters which leads to black hole - white dwarf binaries \citep{2010ApJ...717..948I}. \cite{2019ApJ...881...75K} analyze the number of such events in their simulations and predict about 1 BH-WD binary per GC per Gyr. 
Given a typical lifetime of $10^5$ years for the super-Eddington phase of such objects \citep{2017MNRAS.467.3556B} and that about $2 \times 10^4$ GCs have been searched for such objects \citep[and references therein]{Dage21,2023MNRAS.524.3662N}, suggests these models are consistent at least in a very broad way with the data.

Finally, the properties of the potential second companions to UCXBs would have to be particularly pronounced in RZ2109 due to its high mass transfer rate. Any inferences about the properties of such triple systems and their effects on the long-term evolution of UCXBs, see  e.g., \citet{Toonen2016,Toonen2020}, will be crucial in estimating the expected number of foreground sources for LISA \citep{2023LRR....26....2A}.

\section*{Acknowledgements}
 The authors thank Gilles Bergond, Tom Maccarone, Will Clarkson and Alicia Rouco Escorial for helpful discussion, and the referee for helpful feedback that improved the manuscript.  KCD and DH acknowledge funding from the Natural Sciences and Engineering Research Council of Canada (NSERC), and the Canada Research Chairs (CRC) program. KCD acknowledges fellowship funding from Fonds de Recherche du Qu\'ebec $-$ Nature et Technologies, Bourses de recherche postdoctorale B3X no. 319864. KCD acknowledges support for this work
provided by NASA through the NASA Hubble Fellowship grant
HST-HF2-51528 awarded by the Space Telescope Science Institute, which is operated by the Association of Universities for Research in Astronomy, Inc., for NASA, under contract NAS5–26555.
This work made use of Astropy:\footnote{http://www.astropy.org} a community-developed core Python package and an ecosystem of tools and resources for astronomy \citep{astropy:2013, astropy:2018, astropy:2022}.
S.N. acknowledges the partial support from NASA ATP 80NSSC20K0505 and from NSF-AST 2206428 grant as well as thanks Howard and Astrid Preston for their generous support.
\section*{Data Availability}
All of the data is publicly available and can be downloaded from HEASARC, the Chandra data archive, or the NOAO archive. 

\bibliographystyle{mnras}
\bibliography{example} 

\appendix
\section{NUV detections and upperlimits of all globular clusters}
Below we list the NUV detections, marginal detections and upper-limits. 

\begin{table*}
\caption{Globular clusters with NUV detections. The source marked with $\dagger$ is RZ2109.}
\begin{tabular}{rrrrrrr}
 \hline
RA & Dec & UVW2 Mag &  Err &  Sigma &  V Mag & Err\\ \hline  \hline

  12:29:09	&+07:58:41.02&   22.5 &     0.4 &  3.1 &20.5 &0.7\\
 12:29:23	&+07:55:40.91 &22.7 &     0.3 &   3.2 &  20.5 &0.7\\
 12:29:23	&+07:49:01.49 &22.7 &     0.3 &    3.5 & 20.5 &0.6\\
 12:29:26	&+07:58:12.79 &  22.6 &     0.3 &   3.6 &20.0 &1.3\\
12:29:34	&+07:51:29.30  &22.6 &     0.3 &   3.5 &  21.1 &0.6\\
12:29:40	&+07:58:03.50  &  22.3 &     0.3 &4.2 &   20.1 &0.7\\
 12:29:40	&+07:53:33.29 $\dagger$  & 22.0 &     0.2 &   5.6 & 21.0 &0.7\\
12:29:53	&+07:55:59.92   & 22.5 &     0.3 & 3.7 &  19.7 &1.4\\
   12:30:04	&+07:53:48.30 &22.6 &     0.3 &   3.2 & 20.8 &0.7\\
  12:30:11	&+07:53:45.38  &22.4 &     0.3 & 3.2 &  20.2 &0.7\\ \hline

\end{tabular}
\end{table*}
\begin{table*}
\caption{Marginal detections}
\begin{tabular}{rrrrrrr}
 \hline
RA & Dec & UVW2 Mag &  Err &  Sigma &  V Mag & Err\\ \hline  \hline

 12:29:21	&+07:58:27.80&23.0 &   0.4 &   2.6 &  21.5& 0.5  \\
12:30:04	&+07:51:07.70 &23.0 &   0.4 &   2.7 &  21.2& 0.6\\
 12:29:27	&+07:53:21.59&22.9 &   0.4 &   2.9 &  21.3 &0.9\\
12:29:31	&+07:54:19.69 &23.0 &   0.4 &   2.5 &  21.2 & 0.7\\
12:29:31	& +07:51:12.10 &23.3 &   0.5 &   2.0 &  20.8 &0.5\\
12:29:38	&+07:52:54.41 &23.2 &   0.5 &   2.1 &  20.9 &0.7\\
 12:29:44	&+07:51:52.81&22.9 &   0.4 &   2.8 &  21.3 &0.8\\
 12:29:45&	+07:55:45.80&22.8 &   0.4 &   2.7 &  20.3 &0.6\\
12:29:52	&+07:48:09.40 &22.9 &   0.4 &   2.8 &  21.6 &0.8\\
 12:29:52	&+08:00:00.39&22.4 &   0.4 &   2.6 &  19.8 &1.9\\
 12:29:53	&+08:00:26.60&22.6 &   0.4 &   2.5 &  20.0 & 0.9\\
12:29:58	&+07:54:00.79 &22.8 &   0.4 &   2.8 &  21.2 &0.6\\
12:30:00	&+07:52:07.10 &22.9 &   0.4 &   2.8 &  21.2 &0.6\\
 12:30:02	&+07:52:43.72&23.2 &   0.5 &   2.0 &  21.6 & 0.7\\
12:30:04	&+07:51:07.70 &22.9 &   0.4 &   2.6 &  20.8 &0.6\\
 12:30:06	&+07:49:01.49&22.7 &   0.4 &   2.7 &  21.0 &0.6\\
12:30:09	&+07:51:49.39 &23.0 &   0.5 &   2.0 &  21.2 & 0.5\\\hline
 \end{tabular}
\end{table*}

\begin{table*}
\caption{NUV upper limits of sources detected at less than 2 $\sigma$.}
\begin{tabular}{rrrrr}
 \hline
RA & Dec & UVW2  Mag &  V Mag & Err\\ \hline  \hline

12:29:03 &	+07:53:57.19 & $<$ 23.1 &  20.6 & 0.8 \\
12:29:08	& +07:55:28.99&$<$ 23.3 &  21.5 & 0.9\\
12:29:13	&+08:00:33.41& $<$ 22.9 &  21.5 &0.8\\
12:30:14	&+07:53:37.28& $<$ 23.1 &  21.4 & 0.6\\
12:29:15	&+07:53:07.01&$<$ 23.3 &  21.1 &0.7\\
12:29:18	& +07:44:54.31&$<$ 23.0 &  21.1 & 0.7 \\
12:29:18	&+07:57:39.60&$<$ 23.3 &  21.0 & 0.8\\
12:29:22	&+07:53:51.68&$<$ 23.3 &  21.6 & 0.8\\
12:29:28	&+07:51:37.01&$<$ 23.9 &  21.2 &0.5\\
12:29:30	& +07:47:36.71&$<$ 23.3 &  20.6 & 0.7\\
12:29:30	&+07:53:28.79&$<$ 23.3 &  21.2 &0.7\\
12:29:55 &	+07:48:33.91 &$<$ 23.2 &  21.4 &0.6\\
12:29:39	&+07:48:17.39&$<$ 23.2 &  21.2 & 0.6\\
12:29:40	&+07:56:26.09&$<$ 23.2 &  19.8 & 0.9\\
12:29:42	&+08:00:45.79&$<$ 22.8 &  19.8 & 1.6\\
12:29:42	&+08:00:23.80&$<$ 22.8 &  20.4 & 0.5\\
12:29:42 &	+07:49:50.30&$<$ 23.2 &  20.7 &0.8\\
12:29:44	&+07:53:20.11 &$<$ 23.2 &  21.6 &0.8\\
12:29:45	& +08:01:52.79&$<$ 22.9 &  20.2 & 0.8\\
12:29:50	&+07:54:21.10&$<$ 23.2 &  20.5 & 0.7\\
12:29:52	&+07:45:44.39&$<$ 23.0 &  21.7 & 0.7\\
12:29:35	&+07:54:14.80&$<$ 23.2 &  21.4 & 0.7\\
12:29:53	&+07:55:05.81&$<$ 23.2 &  20.3 & 0.6\\
12:29:53	&+07:51:18.90&$<$ 23.2 &  21.4 & 0.7\\
12:29:56	& +07:53:54.49&$<$ 23.2 &  21.2 & 0.9\\
12:29:56	&+07:52:45.80 &$<$ 23.3 &  21.2 & 0.7\\
12:30:01	&+07:55:23.92&$<$ 22.8 &  19.6 & 1.5\\
12:30:04	&+07:49:01.20&$<$ 23.2 &  21.2 &0.8\\
12:30:05	&+07:50:07.69&$<$ 23.2 &  21.1 & 0.8\\
12:30:05	&+07:53:49.99&$<$ 23.1 &  20.9 &0.7\\
12:29:14	&+07:59:18.60&$<$ 23.0 &  21.4 & 0.7\\
12:30:10	&+07:50:00.82&$<$ 22.9 &  21.3 & 0.8\\  \hline
\end{tabular}
\end{table*}






\bsp	
\label{lastpage}
\end{document}